\documentclass[aps,pre,groupedaddress,11pt]{revtex4-1}
\usepackage{amsmath,amssymb,amsfonts,subfigure,fancybox,txfonts,bm,color,wrapfig}
\usepackage{siunitx,numprint}
\usepackage[dvipdfmx]{graphicx}
\usepackage{bm}      
\usepackage{braket}

\begin{document}

\newcommand{\vc}{\mathbf}
\newcommand{\gvc}[1]{\mbox{\boldmath $#1$}}
\newcommand{\fracd}[2]{\frac{\displaystyle #1}{\displaystyle #2}}
\newcommand{\ave}[1]{\left< #1 \right>}
\newcommand{\red}[1]{\textcolor[named]{Red}{#1}}
\newcommand{\blue}[1]{\textcolor[named]{Blue}{#1}}
\newcommand{\green}[1]{\textcolor[rgb]{0,0.6,0}{#1}}
\newcommand{\del}[3] {\frac{\partial^{#3} #1}{\partial #2^{#3}}}
\newcommand{\dev}[3]{\frac{\text{d}^{#3} #1}{\text{d}#2^{#3}}}
\newcommand{\pdev}[3]{{\text{d}^{#3} #1}/{\text{d}#2^{#3}}}
\newcommand{\pdel}[3]{{\partial^{#3} #1}/{\partial #2^{#3}}}
\newcommand{\intd}[1]{\text{d} {#1}}
\newcommand{\emf}[1]{{\gtfamily \bfseries #1}}
\newcommand{\subti}[1]{\begin{itemize} \item \emf{ #1} \end{itemize}}

\newcommand{\Real}{\operatorname{Re}}
\newcommand{\Imag}{\operatorname{Im}}

\newcommand{\am}{{\bm a}}
\newcommand{\bb}{{\bm b}}
\newcommand{\ff}{{\bm f}}
\newcommand{\pp}{{\bm p}}
\newcommand{\rr}{{\bm r}}
\newcommand{\sm}{{\bm s}}
\newcommand{\tm}{{\bm t}}
\newcommand{\uu}{{\bm u}}
\newcommand{\ww}{{\bm w}}
\newcommand{\xx}{{\bm x}}
\newcommand{\yy}{{\bm y}}
\newcommand{\zz}{{\bm z}}
\newcommand{\Model}{{\mathbb M}}
\newcommand{\RR}{{\mathbb R}}
\newcommand{\NN}{{\mathbb N}}
\newcommand{\oomega}{\mbox{\boldmath $\omega$}}
\newcommand{\WW}{{\bm W}}
\newcommand{\EE}{\mbox{\boldmath $E$}}
\newcommand{\FF}{\mbox{\boldmath $F$}}
\newcommand{\KK}{\mbox{\boldmath $$K$}}
\newcommand{\GG}{\mbox{\boldmath $G$}}
\newcommand{\tr}{\mathrm{T}}
\newcommand{\CC}{\mbox{$\hat{C}$}}
\newcommand{\II}{\mbox{$\hat{I}$}}
\newcommand{\HH}{\mbox{$\hat{H}$}}
\newcommand{\MM}{\mbox{$\hat{M}$}}
\newcommand{\Am}{\mbox{$\hat{A}$}}
\newcommand{\PP}{\mbox{$\hat{P}$}}
\newcommand{\QQ}{\mbox{$\hat{Q}$}}

\title{
Ultrafast single-channel machine vision based on neuro-inspired photonic computing\\
} 

\author{Tomoya Yamaguchi$^{1}$}
\author{Kohei Arai$^{1}$}
\author{Tomoaki Niiyama$^{2}$}
\author{Atsushi Uchida$^{3}$}
\author{Satoshi Sunada$^{2,4}$}
\email{sunada@se.kanazawa-u.ac.jp}
\affiliation{
$^{1}$Graduate School of Natural Science and Technology, Kanazawa University,
Kakuma-machi Kanazawa, Ishikawa 920-1192, Japan\\
$^{2}$Faculty of Mechanical Engineering, Institute of Science and
Engineering, Kanazawa University,
Kakuma-machi Kanazawa, Ishikawa 920-1192, Japan\\
$^{3}$Department of Information and Computer Sciences, Saitama University,
255 Shimo-Okubo, Sakura-ku, Saitama City, Saitama, 338-8570, Japan.\\
$^{4}$Japan Science and Technology Agency (JST), PRESTO, 4-1-8 Honcho,
 Kawaguchi, Saitama 332-0012, Japan\\
}


\begin{abstract}
High-speed machine vision is increasing its importance in both scientific and technological applications. Neuro-inspired photonic computing is a promising approach to speed-up machine vision processing with ultralow latency. However, the processing rate is fundamentally limited by the low frame rate of image sensors, typically operating at tens of hertz. Here, we propose an image-sensor-free machine vision framework, which optically processes real-world visual information with only a single input channel, based on a random temporal encoding technique. This approach allows for compressive acquisitions of visual information with a single channel at gigahertz rates, outperforming conventional approaches, and enables its direct photonic processing using a photonic reservoir computer in a time domain. We experimentally demonstrate that the proposed approach is capable of high-speed image recognition and anomaly detection, and furthermore, it can be used for high-speed imaging. The proposed approach is multipurpose and can be extended for a wide range of applications, including tracking, controlling, and capturing sub-nanosecond phenomena.
\end{abstract}

\maketitle 

\section*{INTRODUCTION \label{sec1}}
The rapid advances in information technologies, including machine learning, have resulted in ever-increasing needs for ultrafast and energy-efficient computing. 
The present computing technologies are based on digital electronics with von Neumann architecture featured by physical separation of information processing and storing.
However, recent advances in machine learning have posed considerable challenges for electronic digital hardware in terms of computing speed and power consumption \cite{Furber_2016,arXiv200705558}.
Besides, transistor scaling in digital electronics is predicted to be unsustainable
\cite{Nahmias:18,Vetter2017ArchitecturesFT}. 
Consequently, the developments of novel computing hardware and concepts are gaining importance \cite{RevModPhys.91.045002,Merolla668,Grollier:2020aa,Markovic:2020aa,9049105,Shastri:2021aa,TANAKA2019100,PhysRevApplied.15.034092,Wright:2022aa,Nakajima:2022aa,Science.abq8271_Delocalized,Sciadv.aay6946_wavephysics}.

The computing using photons as information carriers has attracted considerable attention owing to recent developments of photonic integration and optical communication technologies \cite{9049105,Shastri:2021aa,Tait:2017aa,Lin1004,9064516,Feldmann:2021aa,Xu:2021aa,Kitayama:2019,Shen:2017aa}.
Recent studies have revealed the potential to overcome major bottlenecks in electronic computing, suggesting the achievement of ultrahigh-speed computing with low energy consumption \cite{Feldmann:2021aa,PhysRevX.9.021032,Wang:2022aa}. 
Photonic computing substrates have been predominantly used to process optical analogue signals and play an essential role in the interface between the physical world and digital domain \cite{Kitayama:2019,Wang_arXiv:2207.14293}.
Such photonic approaches are promising to accelerate the pre-processing of signals from sensing units and may enable off-loading heavy computation in electronic post-processing units. 

However, when photonic processing units are used to process signals acquired by sensing devices, 
the processing speed of the entire processing system may be essentially limited by 
data acquisitions in sensing devices and their transfer to the processing units.
This gets serious when image sensors with many pixels are used, where 
spatial information acquired by an image sensor is converted into an electrical domain with a digital format, and huge amounts of memory are required for data storage.
The electrical domain conversion and memory access of large amounts of data are significant bottlenecks that hinder the speeding up of image processing (Fig.~\ref{fig1}A). 
Meanwhile, all information is not always required for the end use of images in machine vision applications.
Compressed data can be directly processed, instead of processing all visual information, in many applications requiring real-time and high-speed control, such as flow cytometry \cite{Ota_Science2018} and target tracking \cite{Yang:22,Zha:22}, 
because natural images can be essentially compressed without information loss \cite{Gibson:20,4472247}. 
The image-free processing can eliminate redundant image acquisition and allows for higher data efficiency, combined with machine learning, such as neural networks (NNs) \cite{4472247,8918920}. 

Photonic hardware has great potential for image processing.
Many photonic or optoelectronic NN processors for image classifications have been well developed \cite{Ashtiani:2022aa,Wang_arXiv:2207.14293,Shi:2022aa,Chang:2018aa,Mennel:2020ab,Lin1004,Antonik:2019aa}; however, most of them require implementing many pixels for high-resolution visual data acquisition and can put a burden on computational resources.
Thus, they face limitations in processing of large amounts of visual information. 
In addition, photonic NN processors incur a significant training cost, which results in the difficulty of in-situ training in edge computing devices and becomes critical under concept drifts \cite{9999257}.   

Reservoir computing (RC), originally known as echo state network \cite{Jaeger78} or liquid state machine \cite{Maass:2002}, is a neuro-inspired computing paradigm based on the responses of dynamical systems driven by inputs \cite{VERSTRAETEN2007391}. 
RC basically uses a randomly connected network with fixed weights, the readout weights are only trained with a simple training method, such as ridge regression. 
Despite the simple training, RC can achieve excellent inference performance in time-series processing. 
More importantly, it can easily be implemented in physical substrates, including photonic ones \cite{TANAKA2019100}. 
Online training and significant reduction of training cost based on transfer learning are also possible \cite{9000710,PhysRevE.102.043301}. 
Photonic RC can exhibit massive scalable operation and is promising as edge computing devices for processing time-series signals \cite{VanderSande:2017aa,Paquot:2012aa,Brunner:2013aa,Vandoorne:2014aa,Ortin:2015aa,Vinckier:15,PhysRevX.7.011015,Takano:18,Uchida:2020JJAP,8807158,Sunada:21,PhysRevX.10.041037,Harkhoe:20}.

Based on an RC concept, we propose an image-sensor-free photonic approach for ultrafast image recognition and detection. 
In our approach, visual information of physical objects can be compressively acquired with only a {\it single} input channel, in contrast to previous approaches based on multi-pixel-based data acquisition \cite{Ashtiani:2022aa,Wang_arXiv:2207.14293,Shi:2022aa,Chang:2018aa,Mennel:2020ab,Lin1004,Antonik:2019aa}. 
This enables highly scalable processing in a computational resource-efficient way and simple systems without large-scale post-processors for image processing.
The key technique for the compressive single-channel data acquisition is the transformation from the object image into time-domain signals based on an optical random pattern projection, inspired by ghost imaging or single-pixel imaging techniques \cite{Gibson:20,PhysRevA.79.053840}. 
The single-pixel-based processing techniques typically require multiple measurements using different mask patterns and suffer from the low switching rates of the mask patterns, typically ranging between tens of Hz and tens of MHz \cite{Hahamovich:2021aa,Xu:18,Shi:21,Wang:2017ab}. 
Thus, a long time is required to acquire image information.
Meanwhile, our approach allows for the switching of random mask patterns over 25 Gigasamples per seconds (GS/s), which is, at least, a three orders of magnitude higher than conventional approaches,
and enables direct photonic processing of temporally image-encoded signal with our photonic RC processor (Fig.~\ref{fig1}B). 

We experimentally demonstrate a gigahertz-rate image recognition and anomaly detection allowing for detecting and capturing fast-switching phenomena.
With a wavelength multiplexing technique, parallel processing is possible to further accelerate data acquisition.
Furthermore, we show that the proposed temporal image-encoding enables a learning-based image reconstruction.
This imaging technique is different from conventional fast imaging based on broadband pulse lasers \cite{Li:aa,Goda:2009aa,Nakagawa:2014aa}, in the sense that the image-recording time is not limited by the pulse width; thus, a fast phenomenon can be captured for a long time over a few microseconds with a flexible time resolution. 
As such, this study offers a multipurpose high-speed machine vision framework, which addresses a wide range of applications ranging from pattern recognition, anomaly detection, and object tracking, to the imaging of a sub-nanosecond phenomenon. 

\section*{RESULTS}
\subsection*{Basic operation principle}
The proposed system architecture includes a random pattern projector to temporally encode the spatial information of target objects and a photonic RC processor to process image-encoded time-domain signals (Fig.~\ref{fig1}C).
The random pattern projector generates random mask patterns, which are illuminated to the target object. 
The light is reflected from the target, of which an image is denoted by $v(x,y)$, focused by a focusing lens, and directly sent to the photonic RC processor, where $(x,y)$ represents the coordinate on the image plane.  
For the random mask pattern $\mathrm{Mask}(x,y,t)$ on the image plane at time $t$, 
the input light $u(t)$ to the RC processor can be characterized by the spatial integral $\int \mathrm{Mask}(x,y,t)v(x,y)dxdy$, i.e., the spatial information of the target image is encoded as a time-domain signal. 
The reservoir plays a role in mapping of input $u(t)$ into a high-dimensional feature space \cite{TANAKA2019100}, so that the features of $u(t)$ can be separately distributed in the high-dimensional space, resulting in better recognition with simple post-processing. 

Here, let $\xx_r(t)$ and $\phi(\xx_r(t)) \in \RR^{M}$ be the reservoir's internal state vector and observables responding to $u(t)$. 
The observables, $\phi(\xx_r(t))$, are sampled at a sampling time interval $\tau_s$ during the acquisition time $T_N$.
In the same way as previous studies \cite{Sunada:21}, the output vector $\yy(n) \in \RR^{M_y}$ $(n \in \{1,2,\cdots\})$ is given by the observables $\phi(\xx_r(t_j))$, readout weights $\WW(t_j) \in \RR^{M_y\times M}$, and bias $\bb \in \RR^{M_y}$ as $\yy(n) = \sum_{j=0}^{N-1}\WW(t_j)\phi(\xx_r(t_j)) + \bb$ for regression tasks, where $t_j = nT_N + j\tau_s$, $(j \in \{0,1,\cdots, N\})$ and $N = T_N/\tau_s$, whereas $\yy(n) = f(\sum_{j}\WW(t_j)\phi(\xx_r(t_j))+\bb)$ for classification tasks, where $f$ is a softmax function.
The weight matrix $\WW(t)$ can be trained with a training dataset such that the output vector $\yy(n)$ correspond to the target vector $\yy_{\mathrm{tag}}(n)$.
In this scheme, the output vector $\yy(n)$ can be obtained with time interval $T_N$. 

A major advantage of using the RC processor compared to standard recurrent neural networks is its fast learning to access the global minimum of a loss function, resulting in low training costs.
The post-processing can be done with an application-specific circuit or field programmable gate arrays for low-latency operation. 
In this study, we focus on evaluating the capability of the fast data acquisition and preprocessing using the RC processor for the proof-of-concept.

\subsection*{High-speed random pattern projector}
The random pattern projector is based on a high-speed {\it speckle} generator developed in our previous study \cite{Hanawa:22}; it is composed mainly of a laser source, random number generator, phase modulator, and multimode fiber (MMF) (Fig.~\ref{fig1}D). 
When coherent light is input to the MMF, the light is coupled into multiple propagation modes with different phase velocities, and their interference produces a speckle pattern at the end face of the MMF \cite{Rawson:80}.
Since speckles are sensitive to the phase change of the incident light, 
the speckle pattern can be changed by modulating the phase of the incident light.  
In existing studies, optical random speckle patterns have been generated using spatial light modulators (SLMs), including digital micromirror devices (DMDs), at rates of 22 kHz at most \cite{Gibson:20}.
A recent promising study demonstrated the modulation rates of up to 2.4 MHz using mechanically rotating mask patterns \cite{Hahamovich:2021aa}.
On the other hand, the modulation rate in the proposed projector can be over tens of GHz, which can be controlled by the phase modulation rate \cite{Hanawa:22}.

\subsection*{Photonic reservoir computing processor}
\begin{figure}[htbp]
\centering\includegraphics[bb=0 0 792 518,width=17cm]{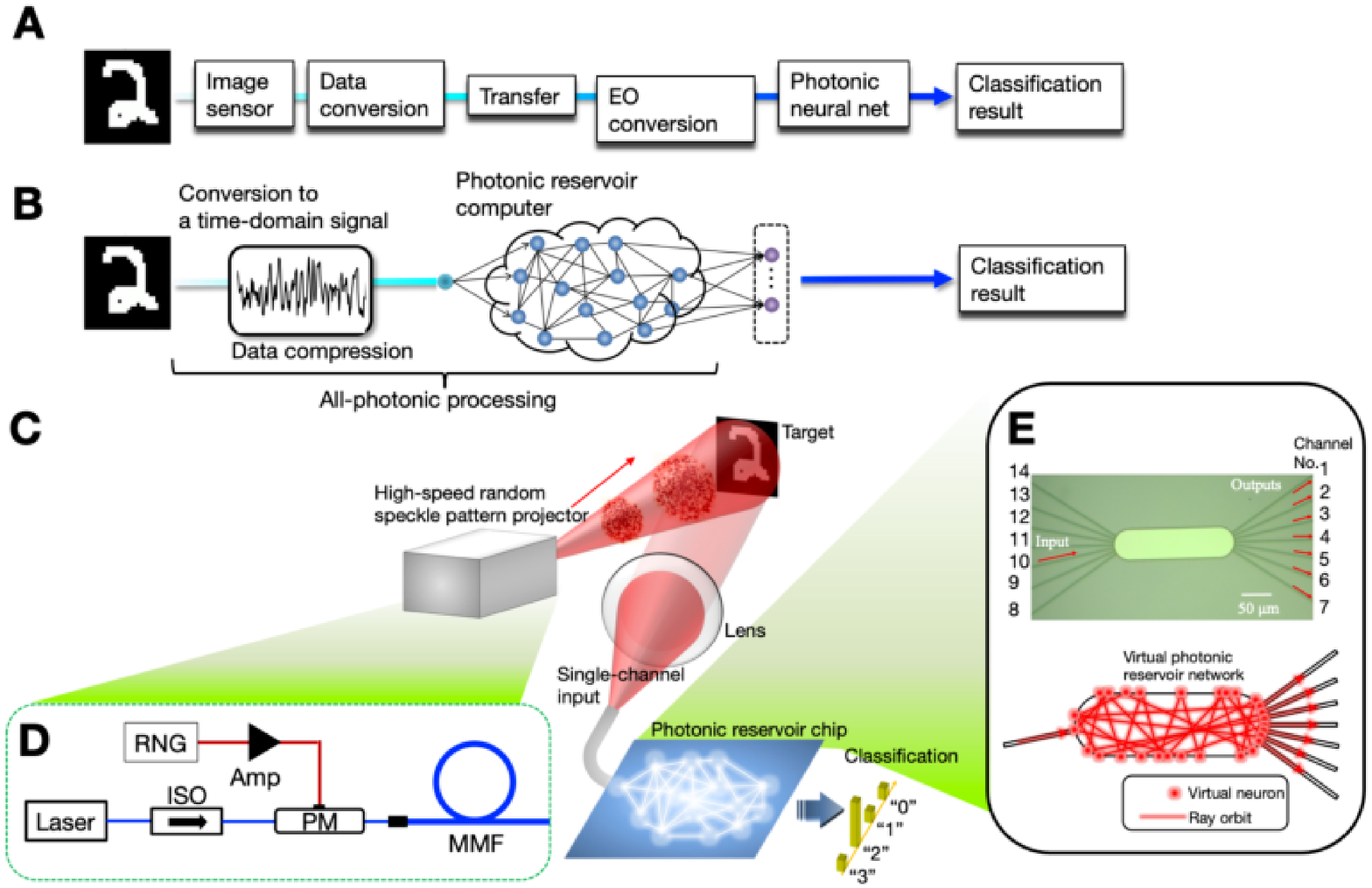}
\caption{\label{fig1}
{\bf Conceptual schematic of photonic machine vision system.}
(A) Conventional approach based on image sensors.
The entire processing rate is limited by the low frame rate of the digital image sensors used for acquiring the visual information of a target object. 
(B) Proposed photonic approach based on a single-channel image acquisition and photonic RC processor. 
(C) Conceptual schematic of the proposed system. 
(D) Setup for high-speed random speckle pattern projector.
RNG: random number generator, ISO: optical isolator, PM: phase modulator, MMF: multimode fiber.
(E) On-chip photonic RC processor based on a stadium-shaped microcavity coupled to input/output waveguide channels.  
The microcavity can form as a virtual random optical network via an internal chaotic multiple scattering from a ray-optic point of view.
In this study, the signal was input from the waveguide channel (No.10) to the reservoir, and the RC output signals were extracted from five output channels (No.2-6) and were used for post-processing. 
}
\end{figure}

The photonic RC processor is based on a stadium-shaped microcavity structure coupled to 14 singlemode waveguides on a silicon chip (Fig.~\ref{fig1}E). 
The microcavity acts as a reservoir, whereas the singlemode waveguides are used as an input/output channels to/from the reservoir. 
A feature of the microcavity is its efficient capability of the optical confinement in a small footprint and the formation of various wave patterns depending on the shape of the microcavity \cite{RevModPhys.87.61}. 
The stadium-shaped cavity is known as a ray-chaotic cavity, inspired by the Bunimovich stadium billiard \cite{Bunimovich:1974aa}.
The wave mixing due to the chaotic nature can form a wave field inside the cavity corresponding to a spatially continuous optical random network in a small footprint (Fig.~\ref{fig1}E), which works as a large-scale reservoir. 
Nonlinearity can be naturally introduced in the intensity detection and phase input processes. 
Numerical results reveal that the stadium-shaped cavity-based RC has high computational performance for tasks requiring nonlinearity, compared to non-chaotic cavity-based RC \cite{Sunada:2019aa}. 
For the capabilities of the temporal signal processing of the stadium-shaped RC chip used in this study, see supplementary text and Figs.~S1-S3, where it is shown that the photonic RC chip can outperform photonic RC systems or a photonic neural network circuit for benchmarking datasets.

\subsection*{Image recognition}
\begin{figure}[htbp]
\centering\includegraphics[bb=0 0 776 538,width=17cm]{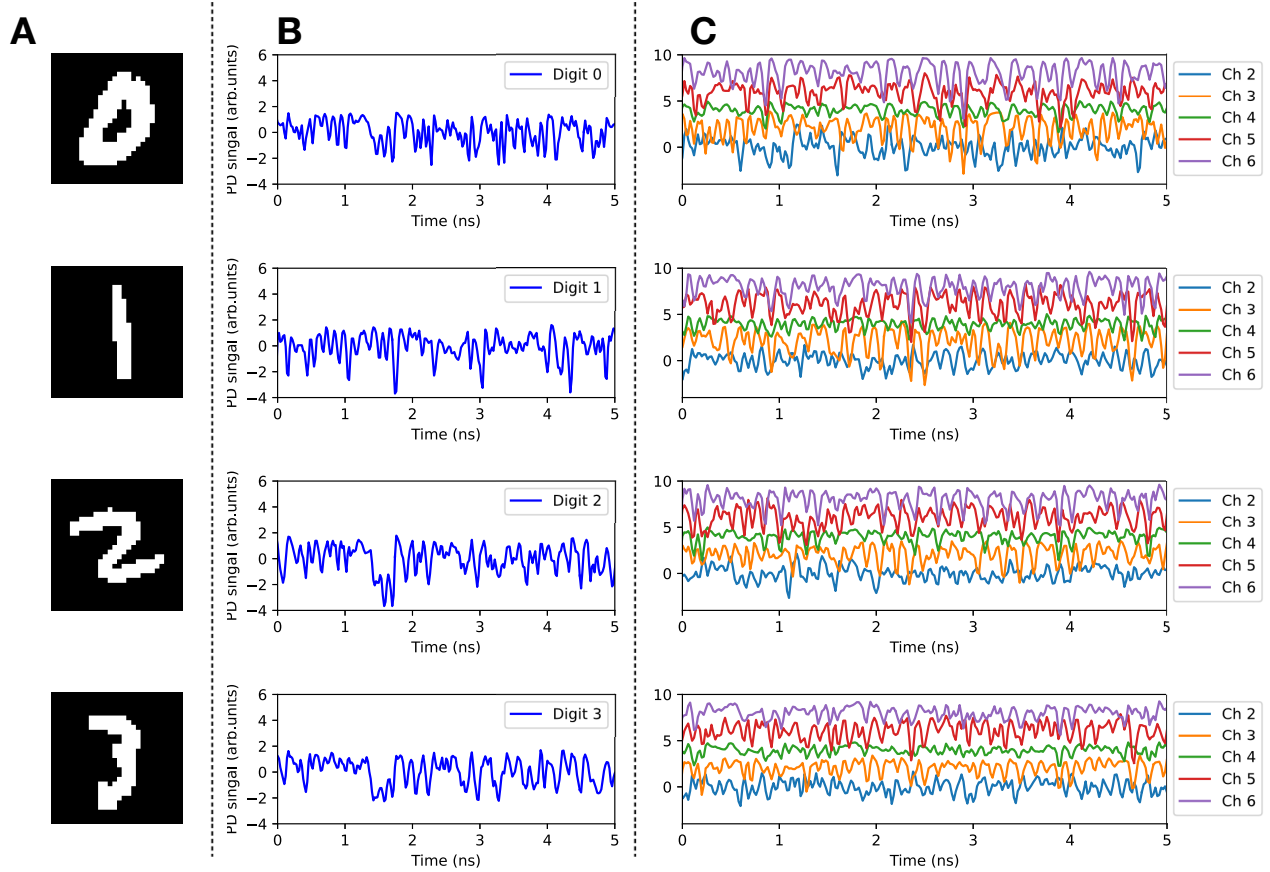}
\caption{\label{fig2}
{\bf Conversion of images into time-domain signals and reservoir outputs.}
(A) Handwritten digit images displayed on the DMD, ``0'', ``1'', ``2'', and ``3'' from the top. 
(B) Image-encoded time-domain signals corresponding to each digit image. 
(C) Outputs from channels (No. 2-6) of the RC processor responding to the time-domain signal.
}
\end{figure}

We evaluated the image recognition performance of the proposed system.
In the experiment, we chose the MNIST handwritten digit images of $28\times 28$ pixels \cite{MNIST_LeCun_url} from ``0'' to ``3'' as the target images and displayed them on a DMD.
The random speckle patterns were generated and projected onto the target at a rate of 25 GS/s. 
The reflected light was input into the photonic RC processor via an optical fiber.
The RC outputs were measured with fast-response photodetectors. 
Figure~\ref{fig2} shows the time evolution of the light intensity reflected from the images of the targets (i.e., an input to the RC) and the corresponding RC outputs from channels No. 2-6 (Fig.~\ref{fig1}E).
The waveforms of the reflected lights strongly depend on the images of the targets, and a variety of spatiotemporal responses of the reservoir outputs are produced. 

\begin{figure}[htbp]
\centering\includegraphics[bb=0 0 640 541,width=15cm]{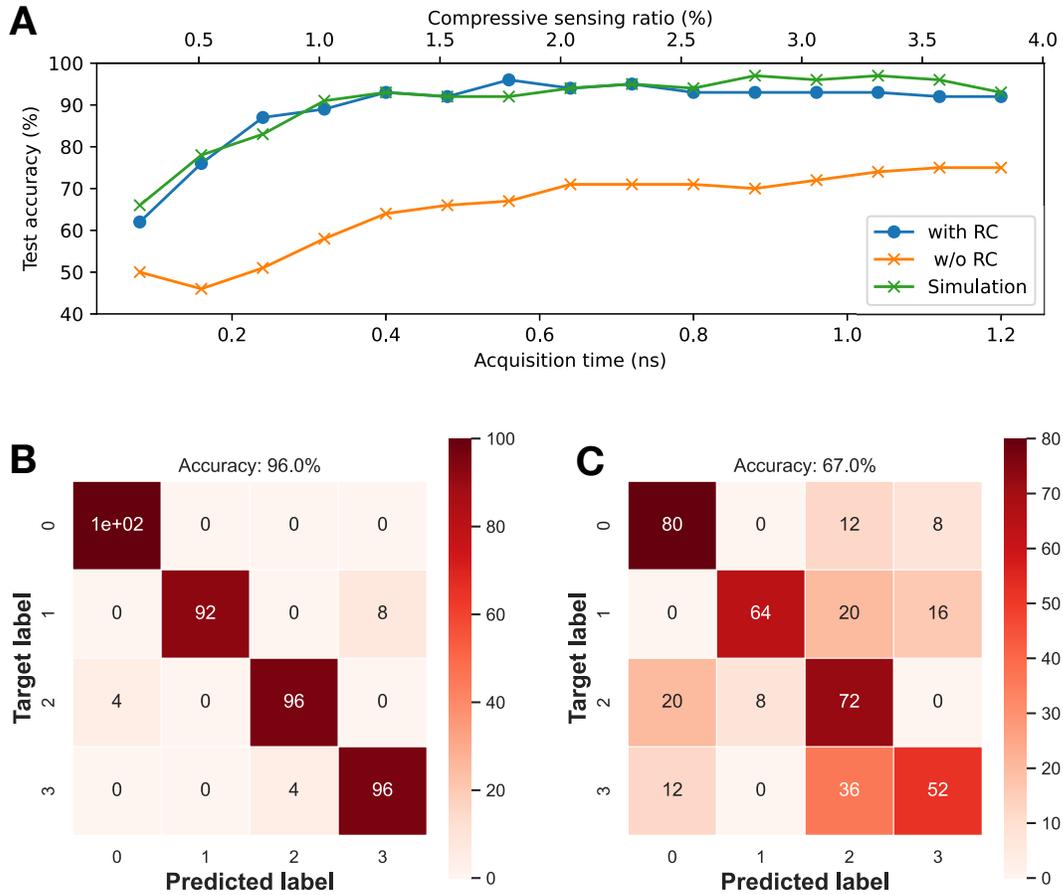}
\caption{\label{fig3}
{\bf Classification results for 4-class handwritten digit image dataset.}
(A) Classification accuracy vs. acquisition time (compressive sensing ratio) for test image samples. 
The accuracy exceeds 90$\%$ for $T_N \ge 0.4$ ns, which corresponds to compressive ratio $C \ge 1.28 \%$.
The performance is better than the performance without the RC processor and is comparable to the accuracy of the numerical neural network with the same number of neurons with $\tanh$ activation functions. 
Confusion matrix for the test image samples in the proposed system (B) with and (C) without the RC processor for $T_N = $ 0.56 ns. 
}
\end{figure}

For the evaluation, we used 1000 samples of digit images from ``0'' to ``3'' and acquired the RC outputs during an acquisition time $T_N$ for each image. 
The prediction outputs, $\yy$, were trained with 900 image samples and tested with 100 image samples.
Figure~\ref{fig3}A shows the classification accuracy for various acquisition times. 
The compressive sensing ratio $C$ for the data acquisition is defined by $N/(28\times 28)$ \cite{8918920}, where $N = T_N/\tau_s$ denotes the number of the data points of the acquired waveform. 
The results show that the accuracy exceeds 90$\%$ for $T_N \ge 0.4$ ns, which corresponds to compressive ratio $C \ge 1.28 \%$, 
revealing the potential of the proposed approach for ultrafast image recognition at sub-nanoseconds with a substantial compression efficiency.
As an example, the confusion matrix for $T_N = 0.56$ ns ($C = 1.78 \%$) is shown in Fig.~\ref{fig3}B.
Most predicted labels were distributed along the diagonal line and matched the true labels. 
For comparison, we performed numerical simulations. 
To mimic the random projection for a digit image (with 28×28 = 784 pixels), 
an $N\times 784$ Gaussian random mask matrix was used. 
As a classifier, we used a neural network with a single fully connected hidden layer and $\tanh$ activation functions.
We confirmed that the classification performance of the proposed system was comparable to that of the numerical neural network (Fig.~\ref{fig3}A).
To gain an insight into the effect of the photonic RC processor, we investigated the classification performance for the system without the RC processor, where the time-domain signal before sending to the RC processor was directly used as an input to a linear classifier. 
The classification performance was much worse than that for the system using the RC processor (Fig.~\ref{fig3}C), suggesting that the spatiotemporal outputs from the RC processor play an important role in the classification. 

We also evaluated the classification performance for a larger and more difficult image dataset. 
The image classification was successfully performed even for such datasets with high compressive efficiency at a nanosecond acquisition time (Fig.~S4). 

\subsection*{Recognizing a microsecond phenomenon}
\begin{figure}[htbp]
\centering\includegraphics[bb=0 0 569 611,width=12cm]{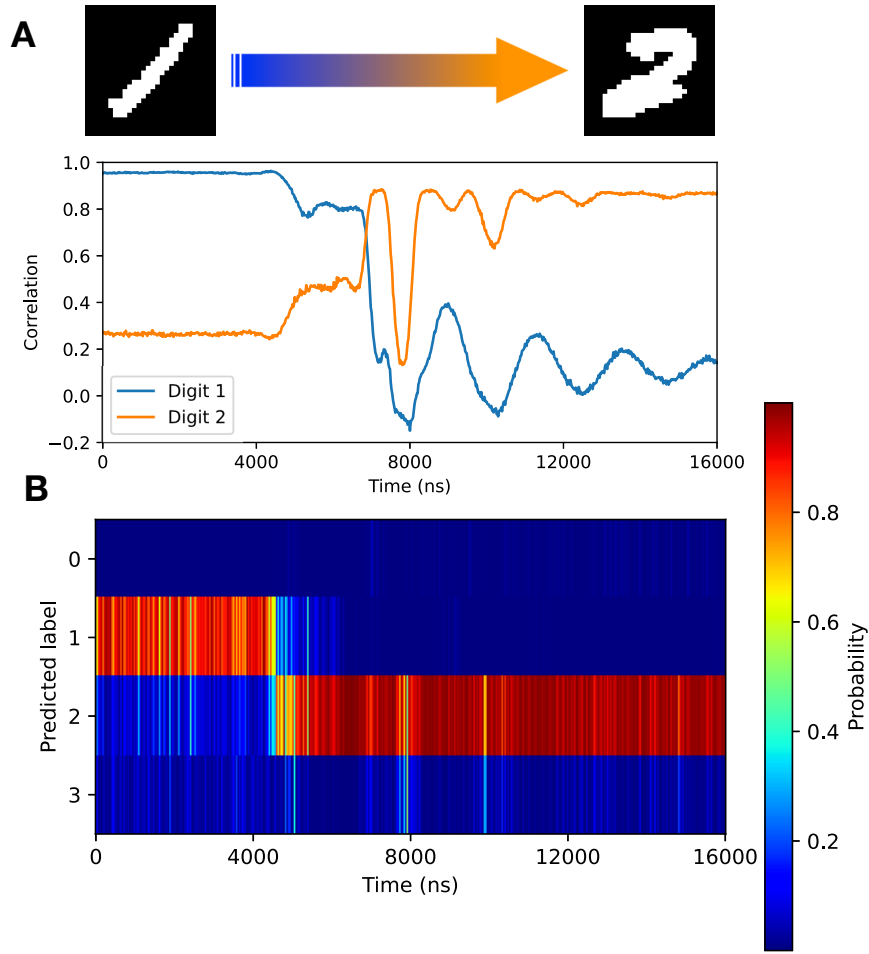}
\caption{\label{fig4}
{\bf Demonstration of dynamic image recognition.}
In this demonstration, the DMD display was switched from digit ``1'' to digit ``2''.
The switching behavior occurred in a few microseconds, and we evaluated the capability of the proposed system for image recognition of the microsecond switching behavior. 
(A) Short-time correlation values for digits ``1'' and ``2'' as a function of time. The correlation analysis reveals that the waveform of the measured time-domain signal changed from that of digit ``1'' to that of digit ``2''. 
The transient behavior of the switching was observed from 4600 ns.
(B) The recognition probability as a function of time. 
}
\end{figure}

To demonstrate the capability of recognizing dynamic scenes, we measured a switching behavior of an image displayed on the DMD from digit ``1'' to digit ``2''.
In the experiment, the laser light was repeatedly phase-modulated with the same pseudorandom signal, and dynamic speckle patterns were repeatedly projected onto the DMD.
The reflected light was directed to our RC processor, and the reservoir outputs were acquired during the acquisition time $T_N = 0.56$ ns to obtain the classification results. 
According to our correlation analysis, the transient of the image switching from digit ``1'' to digit ``2'' occurred from 4600 ns (Fig.~\ref{fig4}A). 
Figure~\ref{fig4}B shows the time-dependence of the classification probability for the transient behavior. 
The result reveals that the image of digit ``1'' was switched to that of digit ``2'' around 4600 ns, and digit ``2'' can be stationarily recognized after the transient. 
The detection of the switching behavior is consistent with the results of our correlation analysis. 
Although the time scale of the DMD display switching is limited to a few microseconds, our system has the potential to recognize and detect faster phenomena. 

\subsection*{Image-free anomaly detection}
\begin{figure}[htbp]
\centering\includegraphics[bb=0 0 515 496,width=12cm]{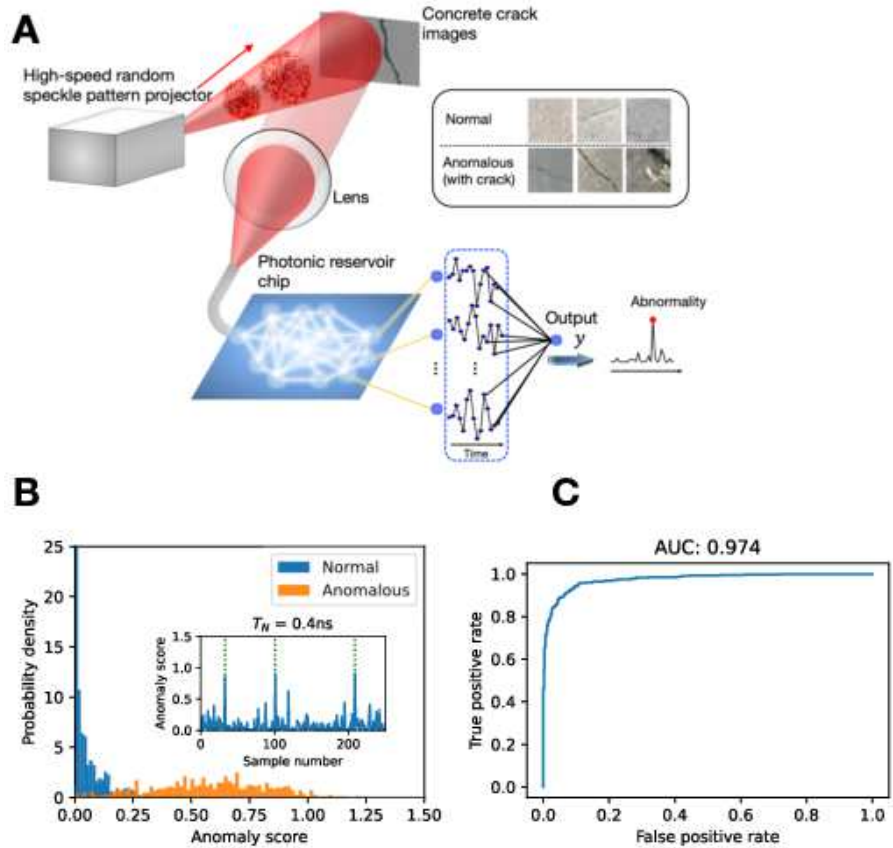}
\caption{\label{fig5}
{\bf Detection of cracks in concrete images.}
(A) Schematic of anomaly detection scheme in images. 
Inset shows examples of normal images (without cracks) and anomalous images (with cracks).
In this experiment, binarized images from a concrete crack dataset were displayed on the DMD. 
The acquisition time was set as $T_N = 0.4$ ns. 
The system was trained using 1500 normal image samples (without cracks) such that the output $y$ corresponds to a nonzero constant value ($\alpha = 1$). 
The squared representation error $|y-\alpha|^2$ was used as an anomaly score. 
(B) The probability densities of anomaly scores from 500 normal images without cracks and 500 anomalous (crack) images. 
The two probability densities are well discriminated. 
The inset shows examples of measured anomaly scores for some sample images. For the display, we set the 3$\%$ of the total samples, represented by the dotted green lines, as crack images. 
(B) Receiver operating characteristic (ROC) curve to illustrate the detection capability of crack images as its discrimination threshold is varied. 
The true positive rate denotes the rate of correctly detecting cracks, whereas the false positive rate denotes the rate of wrongly detecting the absence of cracks. 
The area under curve (AUC) was 0.978. 
}
\end{figure}

Next, we evaluated the feasibility of anomaly detection (Fig.~\ref{fig5}A).
Anomaly detection is an approach to identify an abnormality or rare event from sampled information and is required to operate in realtime as much as possible. 
An anomaly detection from images generally requires heavy computation, which prevents the speeding up of realtime operation.
This problem becomes more serious when considering the implementation in an edge device with limited computational resources.     
Our photonic approach can reduce redundant and unnecessary information of the image data through a compressive transformation into time series data; thus, the required computation for detection can be offloaded. 
This approach also provides merit that image data can be treated in the same manner as the time-series data from the other sensors.
The features of the lightweight computation and low training cost of our approach will enable not only on-device prediction but also on-device learning in edge devices. 

For a demonstration, we used a benchmark dataset of concrete cracks for structural health monitoring and inspections \cite{arxiv.1905.13147_anomaly_images,10.22260_ISARC2018_0094}.
The dataset contains 227$\times$227-pixel concrete images with/without cracks, where each image was taken approximately 1 m away from the surface with a camera directly facing the target \cite{10.22260_ISARC2018_0094}.
The images were displayed on the DMD. 
The system was trained with 1500 normal image samples (without cracks), such that the output $y$ corresponds to a constant value $\alpha = 1$. 
To measure abnormality (images with cracks in this case), an anomaly score was defined as the representation error $(y-\alpha)^2$. 
The score distributes around zero for normal images (without cracks) whereas it has a large outlier when a crack is detected (Fig.~\ref{fig5}B). 
The receiver operating characteristics (ROC) curve is plotted with true positive rates against the false positive rates (Fig.~\ref{fig5}C). 
The area under curve (AUC) was 0.978, which suggests a good measure of separability, considering AUC = 1 for an ideal model. 

\subsection*{High-speed temporal image-encoder for image reconstruction}
\begin{figure}[htbp]
\centering\includegraphics[bb=0 0 1799 1264,width=16cm]{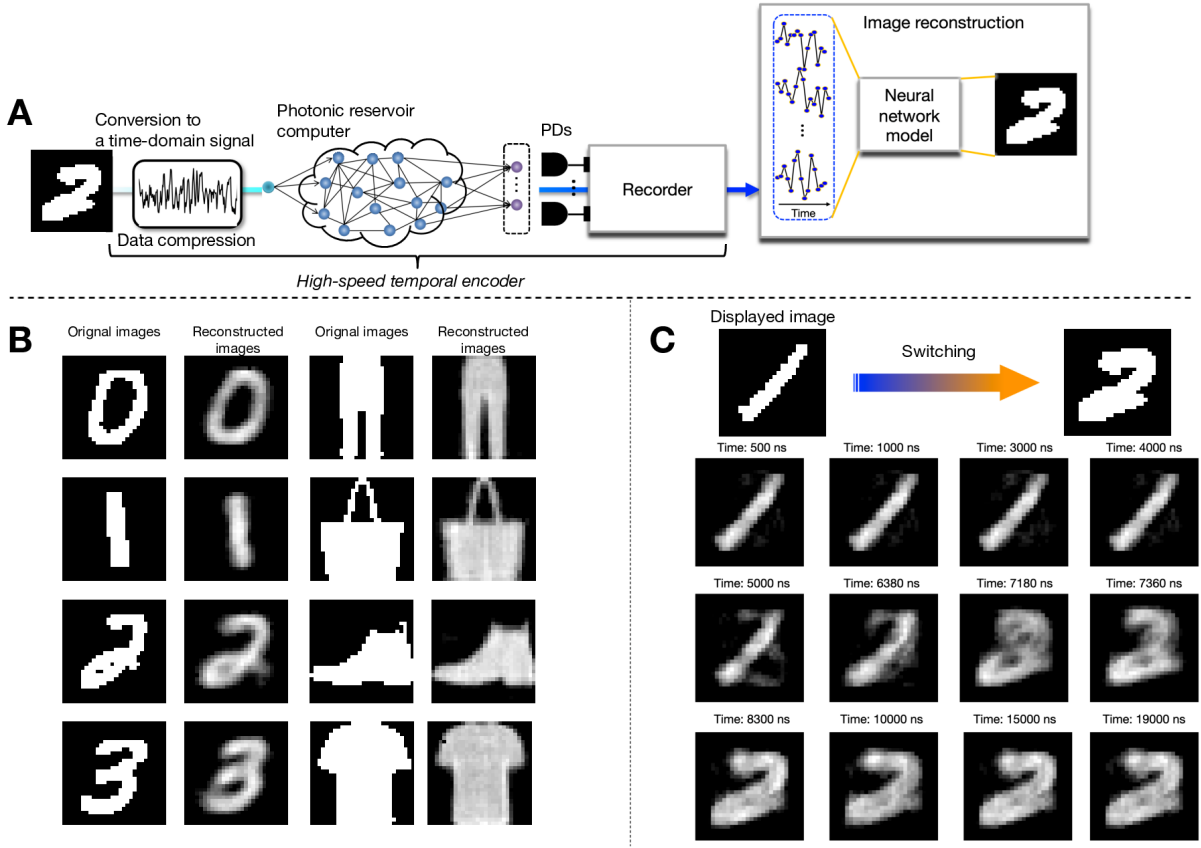}
\caption{\label{fig6}
{\bf High-speed temporal image-encoding and reconstruction.}
(A) Schematic of the high-speed temporal encoder. 
The recorded time-domain signals are used as the inputs to a neural network model for the image reconstruction. 
(B) Examples of the reconstructed images for test samples.  
In the experiment, we used the MNIST handwritten digit image dataset and Fashion-MNIST image dataset and trained the neural network model using 900 image samples for each dataset. 
(C) Reconstructed images during the DMD display switching from digit image ``1'' to ``2''. 
In (B) and (C), the time-domain signals were recorded with the acquisition time $T_N = $ 20 ns. 
}
\end{figure}

As shown above, we illustrated the capability of our system to efficiently process the image information at a remarkable rate.
This leads us to expect that more complex image processing is feasible.
Here, we demonstrate that the proposed system can be used not only as a high-speed recognizer but also as a high-speed imager [Fig.~\ref{fig6}A]. 
A key point is that the reservoir outputs include the image information; thus, the image can be reconstructed from the reservoir outputs using appropriate reconstruction algorithms, e.g., well-developed algorithms for ghost imaging and single-pixel imaging \cite{Gibson:20}. 
However, such algorithms require complete information on the sequences of the projected random mask patterns, which does not apply in our case, because it is difficult to measure the fast spatiotemporal behavior of the random patterns over 10 GHz rates with an image sensor, typically operating at tens of Hz. 
Therefore, we used a neural network model trained with a training dataset to reconstruct a target's image from the measured reservoir outputs (Fig.~\ref{fig6}A). 
It should be noted that realtime processing is not required for this reconstruction purpose.
As a simple proof-of-concept experiment, we used two original datasets containing 4-class handwritten digit images and 4-class Fashion-MNIST images. 
Each image was binarized and displayed on the DMD, and the reservoir outputs were recorded during $T_N = 20$ ns.  
For the image reconstruction, we used a convolutional neural network model trained such that the network output corresponds to the target images. 
We used 900 images for training and 100 images for testing. 
Figure~\ref{fig6}B shows the results of the image reconstruction for some test samples.
The root mean squared errors (RMSE) computed for 100 test images were 0.219 and 0.223 for the MNIST handwritten digit dataset and Fashion-MNIST dataset \cite{arXiv.1708.07747}, respectively.  
Setting a shorter $T_N$-value resulted in an increase in RMSE.
However, the trade-off can be resolved by incorporating the wavelength-division multiplexing (WDM) (Fig.~S5A).
Similar performance was obtained for $T_N \ge 0.8$ ns in the WDM scheme (Fig.~S6).

The proposed temporal image-encoder benefits the observation of a rare event or transient phenomenon.
The proposed approach does not require broadband pulse lasers for temporal encoding of the target images; thus, the recording time is not limited by the pulse width.
Continuous recording for a long time with a controllable time resolution $T_N$ can be achieved. 
To evaluate its feasibility as a primitive experiment, we measured the microsecond switching behavior from digit image ``1'' to digit image ``2'' displayed on the DMD, as demonstrated in Fig.~\ref{fig4}.
In this experiment, the dynamic speckle patterns were repeatedly projected onto the DMD, and the reservoir outputs were acquired with $T_N = 20$ ns. 
Under this condition, the image at each time can be reconstructed with the time resolution of $T_N $.   
As shown in Fig.~\ref{fig6}(C), the switching from digits ``1'' to ``2'' can be observed. 
However, considering that the network was trained only with a 4-class digit images in this study, the reconstructed transient images (shown in the middle Fig.~\ref{fig6}C) may not be correctly captured; the images will be attributed to just the projections of trained digit images. 
Better and more accurate imaging can be achieved by training the reconstruction model with a larger dataset containing independent basis images, e.g., the Hadamard basis patterns \cite{Zhang:17}. 
 
\section*{DISCUSSION}
We experimentally demonstrated a multipurpose, image-free, and ultrafast machine vision approach, which can address image recognition, anomaly detection, and learning-based imaging.  
The proposed approach uses compressive mapping of spatial information into time domain and allows for direct photonic processing with a single input channel to a photonic RC processor, in contrast to conventional approaches replying only on a spatial parallelism of photons. 
Thus, the proposed approach is versatile, has a low training computational cost, and is suitable for the deployment in edge computing devices.
Our gigahertz-rate single-channel data acquisition outperforms conventional approaches. 
Moreover, this is potentially capable of exploiting many advantages revealed in previous studies on ghost imaging or single-pixel imaging, e.g., noise robustness, and image processing with extremely weak intensities.

Further improvement is possible to increase the processing rate. 
One approach is to replace the optical modulator used in this study with the modulator with a wider bandwidth (e.g., over 100 GHz \cite{Wang:2018aa}) because the processing rate is fundamentally limited by the bandwidth of the optical modulator (16 GHz) used for generating speckle dynamics. 
Another approach is to use parallel processing based on another degree of freedom of photons with various multiplexing techniques, such as space-division multiplexing and/or WDM. 
A space-division multiplexing technique can be incorporated using multiple fiber receivers in the proposed approach. 
For WDM, using a multi-wavelength laser (e.g., optical comb) enables the generation of independent speckle patterns in parallel (Fig.~S5A). 
The photonic RC processor also generate output signals in parallel via a demultiplexer. 
Thus, the spatial information of the target images can be encoded and processed in parallel in different domains of time and wavelengths. 
The proposed approach can lead to a significant reduction of the acquisition time of a target image without decreasing the classification accuracy (Fig.~S5B). 

An important future work is to develop a post-processor for an end-to-end fast photonic processor.
One approach will be to deploy a photonic post-processing technique developed as an analog readout in a reservoir computer, which is based on a balanced Mach Zehnder (MZ) modulator and an integrator \cite{Duport:2016aa}, and the multiply-accumulation operation can be achieved in the time-domain.
The additional merit of the analog computation in the time domain is that it can be performed with ultra low energy and a weak signal at a single-photon level can be processed in principle \cite{Science.abq8271_Delocalized}. 
The end-to-end photonic processor will be used for a wide range of applications requiring realtime processing, such as anomaly detection, flow cytometry \cite{Ota_Science2018}, dynamic target tracking and control\cite{Zha:21,SHI2019155,Zha:22}, and potentially for recognizing and detecting unknown ultrafast phenomena. 

Another important future work is to develop a compact random pattern projector. 
In the present experiment, a MMF was used to generate speckle patterns, whereas disordered photonic chips have recently been well developed \cite{Sunada:21,Redding:2013aa}.
Replacing MMF with such disordered chips will enable the all on-chip photonic processor which is capable of being implemented in an edge device. 

We also demonstrated that the proposed approach can be used also as a temporal encoder for high-speed imaging.
In contrast to conventional approaches on high-speed imaging based on using femtosecond pulse lasers  \cite{Li:aa,Goda:2009aa,Nakagawa:2014aa}, the proposed approach is simple, versatile, and has a continuous time recording of a target scene for a long time, that is, recording without dead time (i.e., the time during which the system cannot operate) is possible. 
Furthermore, our approach enables us to capture a wide range of time-scale phenomena by varying the modulation rate and controlling the acquisition time. 
A drawback of our approach is the trade-off between the resolution of acquired images and the acquisition time.
However, incorporating the WDM techniques can resolve the trade-off; the image acquisition can be achieved at shorter time scale with the suppression of degrading the image resolution (Fig.~S6), which can open a novel pathway for the imaging of ultrafast dynamic phenomenon.
\section*{MATERIALS AND METHODS}

\subsection*{Experimental setup}
In our random speckle pattern projector, a narrow linewidth tunable laser (Alnair Labs, TLG-220, linewidth $<$100 kHz, 30 mW) was used as a coherent light source. 
To dynamically generate speckle patterns, the laser light was phase-modulated using a lithium niobate phase modulator (EO Space, PM-5S5-20-PFA-PFA-UV-UL, 16 GHz bandwidth) with a uniformly distributed pseudorandom sequence generated using an arbitrary waveform generator (Tektronix, AWG70002A, 25 GS/s).
The modulated light was directed through a polarization-maintaining single-mode fiber to the MMF, which is a commercially available step-index multimode fiber with a core diameter of 200 $\mu$m, numerical aperture (NA) of 0.39, and length of 20 m. 
The reflected light from the DMD was collected using a focusing lens, coupled to a multimode fiber with a core diameter of 50 $\mu$m, and directed to the photonic RC processor. 
The outputs from the photonic RC processor were measured with photodetectors (New Port, 1554-B, 12 GHz bandwidth). 
To evaluate the performance, the measured signals were digitized using a digital oscilloscope (Tektronix, DPO72504DX, 25 GHz bandwidth) and post-processed using a computer. 

\subsection*{Photonic RC processor}
The RC processor was fabricated on a silicon chip. 
A 220 nm thick silicon layer was etched to form a stadium-shaped microcavity coupled with 14 single-mode waveguides.
The single-mode waveguides were used as the input/output channels.
The stadium was shaped with two semicircles of radius 25 $\mu$m and two parallel segments of length 100 $\mu$m.
The width of the singlemode waveguide was 500 nm. 
A spot-size converter was used to couple the singlemode waveguide and an optical fiber. 

\subsection*{Post processing for image recognition}
The reservoir outputs were detected at a sampling time interval $\tau_s$ during the acquisition time $T_N$.
For the $M$ reservoir outputs with the record length of $N = T_N/\tau_s$, $MN$ features were used as the inputs of the (linear) Softmax classifier. 
The training of the classifier was performed using Python (scikit-learn package). 

\subsection*{Image reconstruction}
In the reconstruction task, we used the reservoir outputs from channel No. 2 to No. 6 ($M = 5$), which were sampled at a sampling time interval of $\tau_s =$ 40 ps. 
In preprocessing, the reservoir outputs were normalized with the means and standard deviations.
The number of the sampled data points for each reservoir output was $N = T_N/\tau_s$, thus, the $MN$ sampled data points were used as input to a neural network model for the image reconstruction.
($T_N$ was set as various values ranging from 0.2 ns to 20 ns.)
In the network model used to obtain the results shown in Fig.~\ref{fig6}B, the full connection network of size $MN\times 200$ was used in the first layer. 
The outputs were sent to the first one-dimensional (1D) CNN layer with 10 kernels of size 3 and the ReLU activation function, followed by batch normalization and max pooling of size 2$\times$2. 
The second 1D CNN layer used a single kernel of size 3 and the ReLU function, followed by batch normalization and max pooling of size 2$\times$2. 
Then, in the fourth and fifth layers, the full connection networks of 50$\times$784 and 784$\times$784 were used to output the 28$\times$28-pixel image.
The network model was trained with $K = $900 image samples to minimize the mean squared error:
$E = (1/K)\sum_{k}^K\sum_{i,j}(I_k(i,j) - I_k^{(target)}(i,j))^2$, where $I_k(i,j)$ and $I_k^{(target)}(i,j)$ are the pixel values of the reconstructed image and target image in the $i$-th row and the $j$-th column for $k$-th sample, respectively, and the model was tested with a separate set of 100 image samples. 

\section*{Funding:}
This work was supported in part by JST PRESTO (JPMJPR19M4), JPSJ KAKENHI (19H00868, 20H04255, 22H05198), and JKA promotion funds from KEIRIN RACE (2022M-208).


\newpage
\section*{Figures}

\clearpage
\renewcommand{\thefigure}{S\arabic{figure}}
\setcounter{figure}{0}
\section*{Supplementary Materials}

\noindent {\bf \Large{Ultrafast single-channel machine vision based on neuro-inspired photonic computing
}}

\noindent {\center Tomoya Yamaguchi$^{1}$, Kohei Arai$^{1}$, Tomoaki Niiyama$^{2}$, Atsushi Uchida$^{3}$, Satoshi Sunada$^{3,4\ast}$}\\

\noindent
$^{1}$Graduate School of Natural Science and Technology, Kanazawa University,
Kakuma-machi Kanazawa, Ishikawa 920-1192, Japan\\
$^{2}$Faculty of Mechanical Engineering, Institute of Science and
Engineering, Kanazawa University,
Kakuma-machi Kanazawa, Ishikawa 920-1192, Japan\\
$^{3}$Department of Information and Computer Sciences, Saitama University,
255 Shimo-Okubo, Sakura-ku, Saitama City, Saitama, 338-8570, Japan.\\
$^{4}$Japan Science and Technology Agency (JST), PRESTO, 4-1-8 Honcho,
 Kawaguchi, Saitama 332-0012, Japan\\

\section{Computational performance of a photonic RC processor}
As a demonstration of processing with the photonic RC chip used in this study, 
we used two standard benchmarks, the Santa Fe time-series prediction task \cite{298828} and vowel recognition task \cite{Deterding1990}. 
\subsection{Chaos time-series prediction}
\begin{figure}[htbp]
\centering\includegraphics[bb=0 0 649 921,width=15cm]{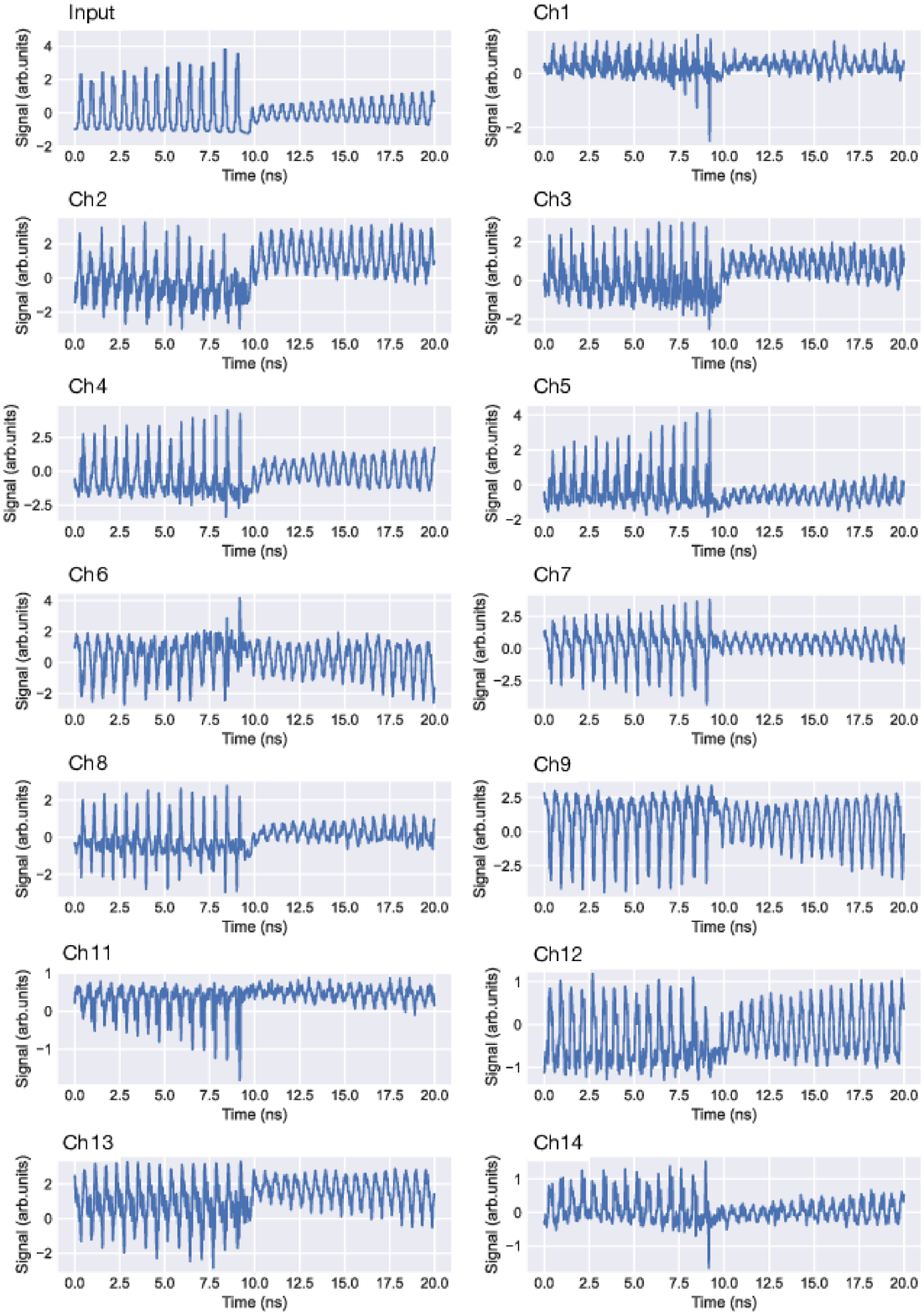}
\caption{\label{suppl_fig1}
{\bf Input chaotic waveform and reservoir outputs.}
Time evolution of each reservoir output $\phi_m(t)$ for channel $m$. 
The input channel was set as the channel No.~10. 
}
\end{figure}

The goal of the Santa Fe time-series prediction task is to predict one step ahead values of the chaotic data generated from a far-infrared laser. 
In this task, the input signal $u(n)$ corresponds to the $n$th sampling point of the chaotic waveform (Fig.~\ref{suppl_fig1}). 
The target $y_{\mathrm{tag}}(n)$ was set to $u(n+1)$ for input $u(n)$. 
To compute the output $y(n)$, the reservoir outputs $\phi_m(t)$ from channel $m$ ($m = 1,2, \cdots, M$), were sampled $N$-times at a sampling time interval of $\tau_s = T_N/N = $ for $n T_N \le t < (n+1) T_N$. 
The output $y(n)$ is given by $MN$-sampled signals as follows:
\begin{eqnarray}
y(n) = 
\sum_{j=0}^{N-1}\sum_{m=0}^M w_{jm}\psi_m(nT_N + j\tau_s),
\end{eqnarray}
where $\phi_0(t) = 1$ is a constant for bias. 
For a training dataset with $N_c$ samples, the weights, $w_{jm}$, were trained such that the following mean squared error ($\mathrm{MSE}$) is minimized: 
\begin{eqnarray}
\mathrm{MSE} = 1/N_c\sum_{n=1}^{N_c} (y(n)-y_{\mathrm{tag}}(n))^2. 
\end{eqnarray}
In this study, we used $N_c = 3000$ step points for training and 1000 step points for testing. 
 
\begin{figure}[htbp]
\centering\includegraphics[bb=0 0 802 324,width=10cm]{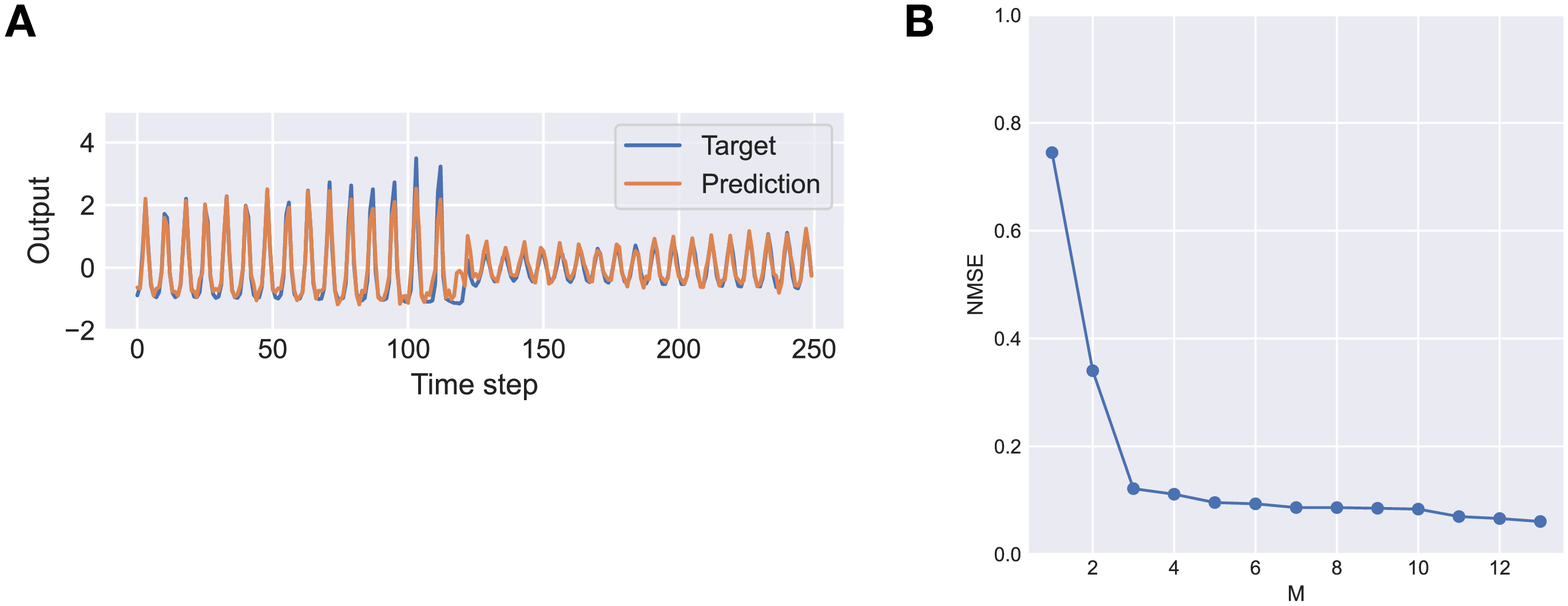}
\caption{\label{suppl_fig2}
{\bf Results for the Santa Fe chaotic time series prediction task.}
(A) Result of the one-step-ahead prediction. The output $y(n)$ was obtained from the reservoir outputs, $\phi_m(t)$, shown in Fig.~\ref{suppl_fig1}. 
For reference, the target $y_{\mathrm{tag}}(n)$ is also shown in this figure. 
(B) The NMSEs as a function of the number of reservoir channels, $M$, used for the prediction. 
}
\end{figure}

Figure~\ref{suppl_fig1} shows the time evolution of the reservoir outputs, $\phi_m(t)$, when the light modulated by a chaotic waveform at a rate of 12.5 GS/s is input from the reservoir channel No. 10 (See Fig.~1E in the main text), demonstrating a variety of dynamical responses to the input chaotic signal $u(t)$.   
The output $y(n)$ was computed for $M = 13$ and $N = 4$, and the prediction performance was measured using the normalized MSE (NMSE). 
The resulting NMSE was approximately 0.08(Fig.~\ref{suppl_fig2}A), which is better than those reported in previous photonic reservoir systems \cite{Takano:18,Brunner:2013aa}. 
Then, we changed the number of the reservoir channels $M$ used for computing $y(n)$ and measured the NMSEs for the prediction task. 
As shown in Fig.~\ref{suppl_fig2}B, the NMSE rapidly decreases as $M$ increases. 
For $M \ge 3$, the NMSEs were less than 0.2.
This result suggests that the number of the detectors and the computational cost in the post-processing can be reduced with moderating the degradation of the prediction. 

\subsection{Vowel recognition}
\begin{figure}[htbp]
\centering\includegraphics[bb=0 0 718 330,width=10cm]{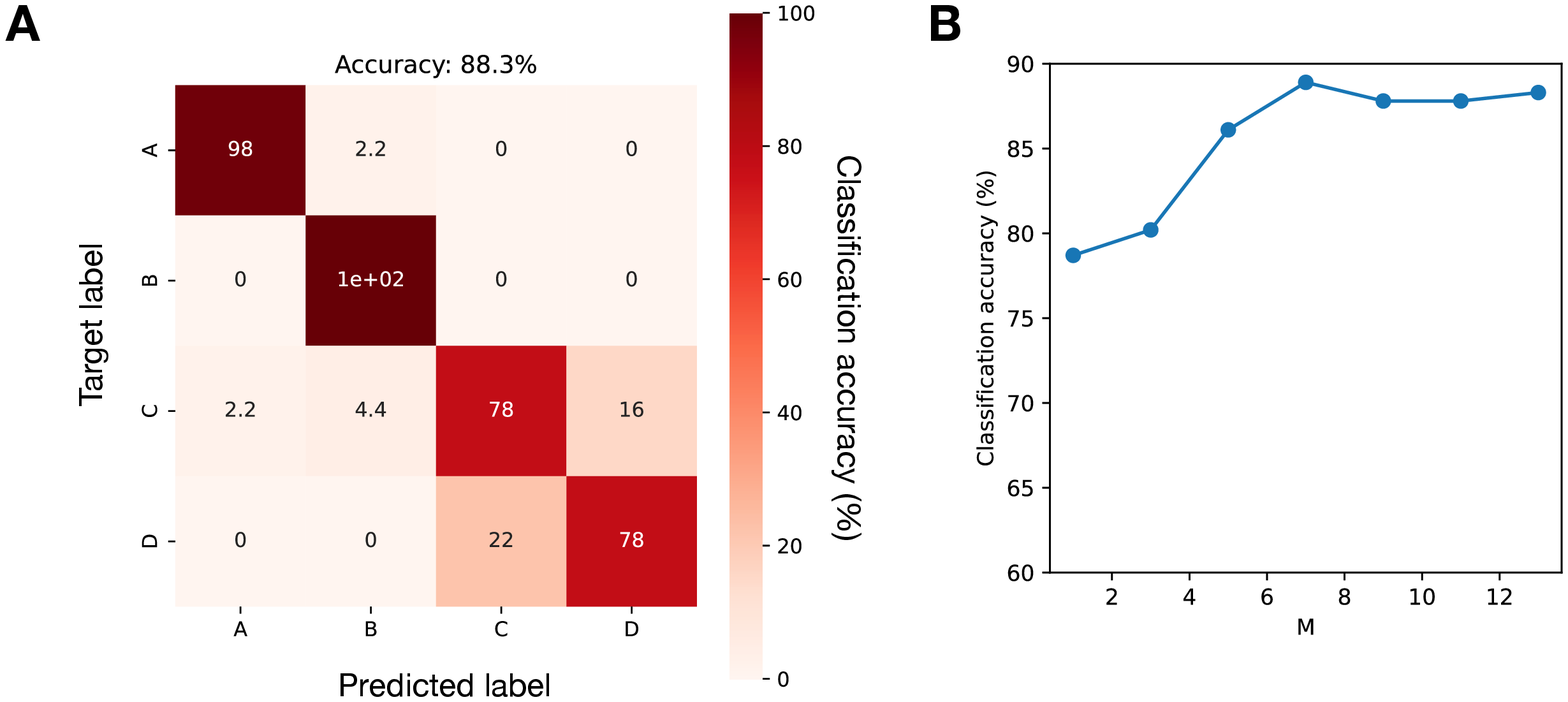}
\caption{\label{suppl_fig3}
{\bf Results for the vowel recognition task.}
(A) Confusion matrix. Spoken vowels, A, B, C, and D, denote ``hid'', ``hEd'', ``hYd'', and ``hOd''. 
The test accuracy was 88.3 $\%$ for 180 samples. 
Misclassification mainly occurred with the confusion between C and D. 
This is because the two vowels are relatively close in a parameter space, as shown in \cite{Shen:2017aa}. 
(B) Classification accuracy as a function of $M$. 
}
\end{figure}

The dataset contains 11 vowel phonemes spoken by 90 different speakers \cite{Deterding1990}.  
The raw data consisted of the steady-state portion of a given vowel utterance, which was low-pass filtered at 4.7kHz and sampled at 10kHz, producing an amplitude signal of approximately 3000 units long. 
The vowel signals were preprocessed to extract 10 long area parameters, which contain the essential information and are used for the recognition task. 
For the training and testing datasets, we used 360 data samples generated by 90 different people speaking four different vowel phonemes. 
We used half of these data samples for the training. 
In the recognition experiment, laser light was phase-modulated using the log area ratio parameters and input to the RC processor. 
The $M = $13 outputs from the RC processor were detected with photodetectors and used for the Logistic regression. 
We evaluated the performance of the vowel recognition with the 180 test samples, and it turned out that the RC processor correctly identified 159/180 samples with an accuracy of 88.3$\%$. 
The confusion matrix is shown in Fig.~\ref{suppl_fig3}. 
A similar evaluation has been performed with an optical circuit, which can compute a vector-matrix multiplication with four channels, and digital processing \cite{Shen:2017aa}. 
The processor achieved an accuracy of 76.7 $\%$. 
For further comparison, we also evaluated the performance of a digital neural network, which consists of a single hidden layer with 100 neurons and ReLU activation functions. 
The neural network was trained with the Adam solver, and achieved 96.1 $\%$ accuracy, which is higher than that of our photonic RC. 
The difference may be attributed to the measurement noise of photonic reservoir outputs. 

We also evaluated the classification accuracies for various $M$ values. 
The accuracy of over 80 $\%$ was obtained for $M \ge 3$.

\section{Fashion MNIST}
\begin{figure}[htbp]
\centering\includegraphics[bb=0 0 1492 1297,width=17cm]{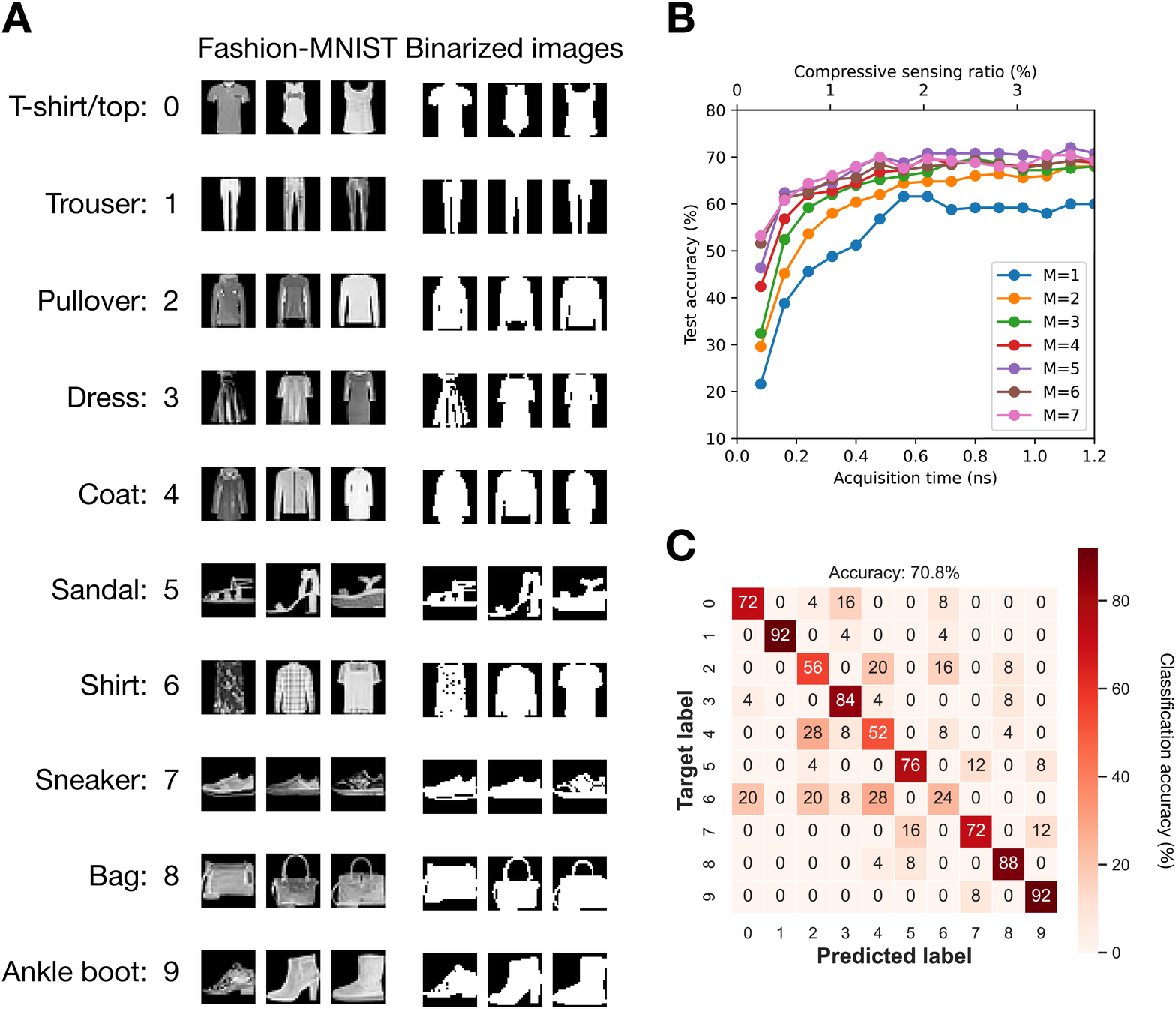}
\caption{\label{suppl_fig4}
{\bf Classification for the Fashion-MNIST image dataset.}
(A) Original fashion-MNIST images and their binarized images. 
The binarized images were displayed on the DMD used in the experiment. 
(B) Test accuracy as a function of acquisition time $T_N$. 
$M$ denotes the number of the reservoir outputs. 
(C) Confusion matrix for the compressive sensing ratio of approximately 3 $\%$ for $T_N = 1$ ns and $M = 5$. 
}
\end{figure}

Fashion-MNIST is a dataset comprising 28$\times$28 grayscale images of fashion products from 10 categories \cite{arXiv.1708.07747}.
For our experiment, the fashion images were binarized to be displayed on DMD(Fig~\ref{suppl_fig4}A).
We used a subset of 2500 sample images from the original dataset for evaluating the recognition performance of the proposed photonic approach. 
Figure~\ref{suppl_fig4}B shows the classification accuracy for 250 test images. 
The experiment was conducted under the same conditions as mentioned in the main text. 
Figure~\ref{suppl_fig4}B shows the test accuracy as a function of the acquisition time $T_N$ or the compressive sensing ratio for various $M$-values. 
The test accuracy increases with the increase in the acquisition time $T_N$. 
The test accuracy was 70.8 $\%$ for $T_N = $ 1 ns, where
the compression sensing ratio was approximately 3 $\%$. 
The misclassifications mainly occurred for the prediction of label 6(Shirt). %
The images of the``6'' were confused as those of 0(T-shirt/top), 2(Pullover), and 4(Coat). 
This is due to the binarization of these fashion images, as shown in Fig.~\ref{suppl_fig4}A. 
For reference, numerical simulations were performed with a neural network with the same number of neurons and achieved 81 $\%$ accuracy. 
The experimental result was worse than the numerical result. 
This may be attributed to the measurement noise of the reservoir outputs caused using a 8-bit digital oscilloscope.
Better results are expected when a larger training dataset is used. 

\section{Wavelength-division multiplexing}

\begin{figure}[htbp]
\centering\includegraphics[bb=0 0 900 852,width=15cm]{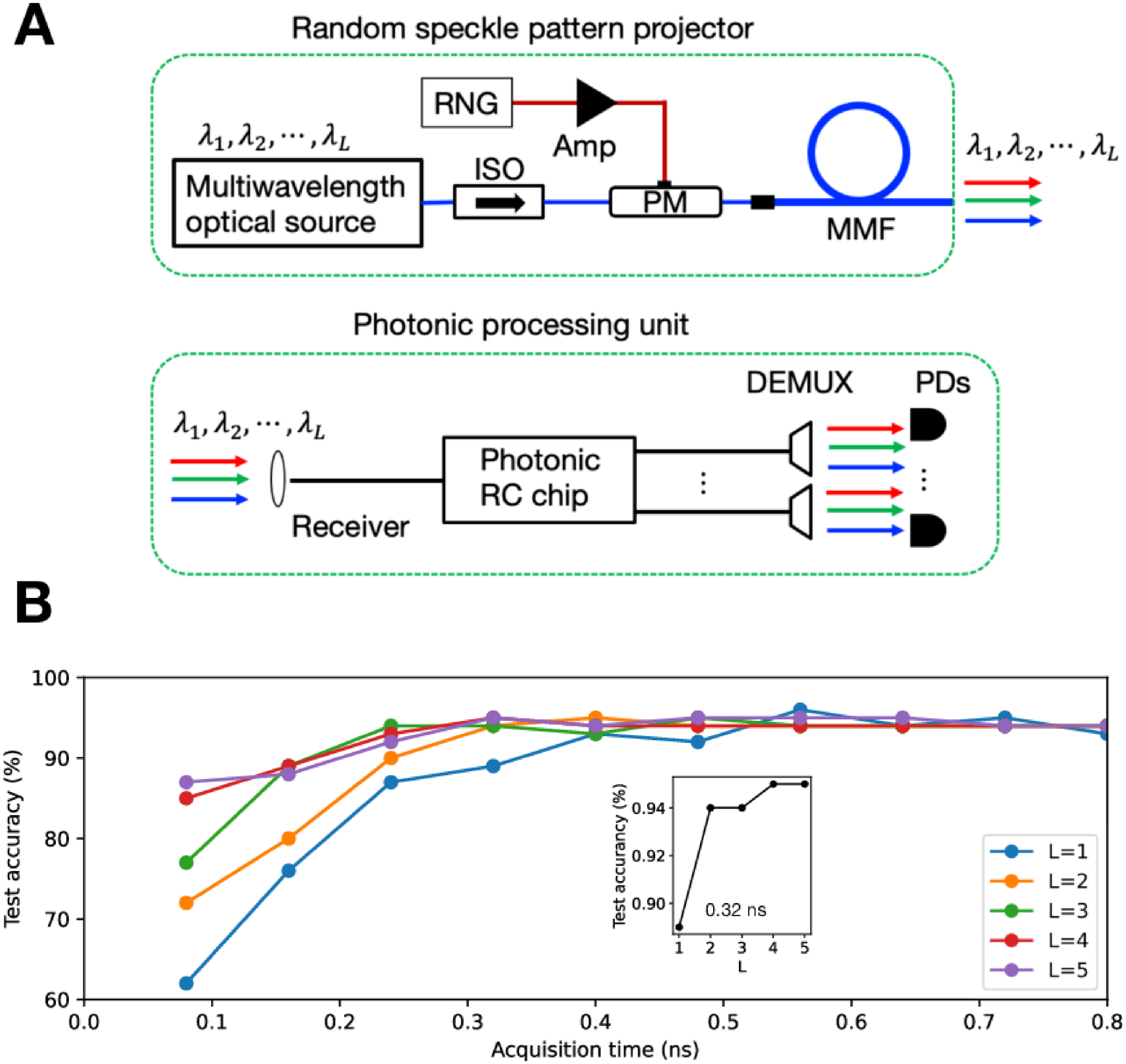}
\caption{\label{suppl_fig5}
{\bf Application of wavelength-division multiplexing and its classification performance.}
(A) The wavelength-division multiplexed system.
A multiwavelength optical source operates at $L$ wavelengths. 
ISO: optical isolator, RNG: random number generator, MMF: multimode fiber, DEMUX: demultiplexer. 
(B) Classification accuracy as a function of acquisition time $T_N$ for different numbers of wavelengths used in the experiment $L$.
As $L$ increases, the test accuracy can be improved for a shorter $T_N$.  
}
\end{figure}

Figure~\ref{suppl_fig5}A shows the proposed approach incorporating wavelength-division multiplexing (WDM). 
A multiwavelength optical source (e.g., optical comb) can be used for the parallel generation of random speckle patterns with different wavelengths.
The optical signals from a target object are sent to a photonic chip, where the signals are processed in parallel.
Then, the reservoir outputs are demultiplexed and sent to subsequent post-processing units.
The parallelism in this scheme is based on the independence between speckle patterns with different wavelengths. 
It is well known that for a long MMF, the speckle patterns are sufficiently decorrelated with a slight wavelength change \cite{Redding:13}.
The wavelength interval for the decorrelation can be estimated as less than 0.02 nm for a 20 m MMF in this experiment.
Therefore, we set the wavelength interval as 1 nm demonstrably to generate independent speckle patterns in this experiment. 

To simply test the effect of the proposed WDM approach for image classifications, we used a tunable laser to produce the laser light with different wavelengths ranging from 1550 nm to 1554 nm at 1 nm and repeatedly acquired the reservoir output signals at each wavelength to mimic the parallel acquisition scheme. 
Figure~\ref{suppl_fig5}B shows the test accuracy for the 4-class MNIST handwritten digit image dataset as a function of the acquisition time $T_N$.  
$L$ is the number of the wavelengths used for the classifications. 
The test accuracy decreases as $T_N$ gets shorter, 
However, the reduction of the accuracy can be moderated as the number of wavelengths $L$ increases. 
For example, at $T_N = 0.32$ ns, the test accuracy decreases to less than 90 $\%$; however, the test accuracy can be increased as $L$ increases.
These results suggest that the proposed approach enables image recognition while suppressing the accuracy reduction even for a shorter $T_N$.

\begin{figure}[htbp]
\centering\includegraphics[bb=0 0 584 1011,width=10cm]{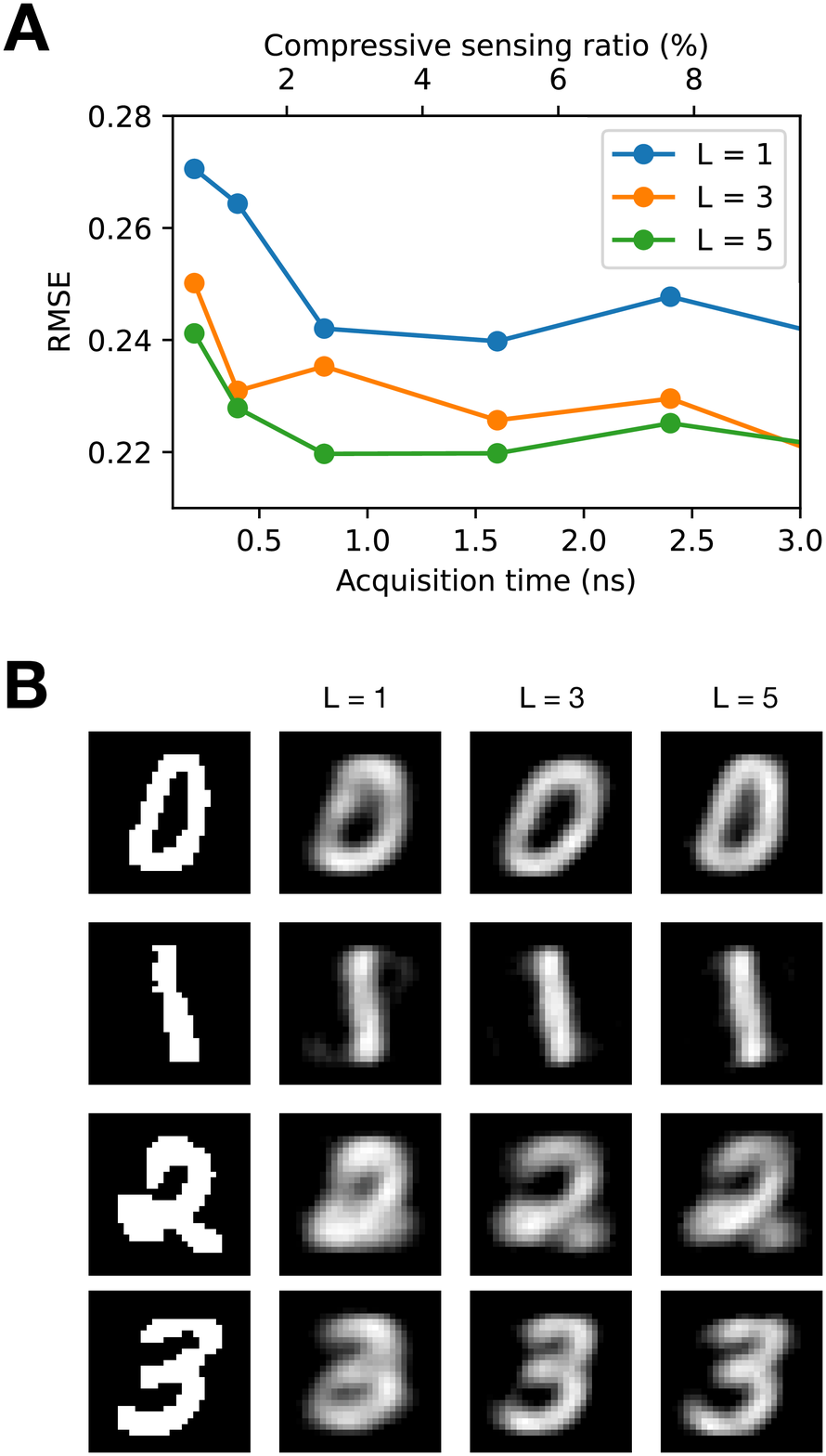}
\caption{\label{suppl_fig6}
{\bf Wavelength-division multiplexed image reconstruction.}
(A) Root mean square error (RMSE) for the image reconstructions as a function of acquisition time $T_N$. 
RMSE decreases as $L$ increases.
(B) Images reconstructed at $T_N = 0.8$ ns for $L =$1, 3, and 5. 
}
\end{figure}

Next, we investigated the effect of the WDM on the image reconstruction. 
In this experiment, we used the same 4-class MNIST handwritten digit image dataset, and the performance was measured using the root mean squared error (RMSE) of the reconstructions for 100 test image samples. 
As shown in Fig.~\ref{suppl_fig6}A, as the acquisition time $T_N$ decreases, the RMSE increases. 
The examples of the reconstructed images at $T_N = 0.8$ ns 
are shown in Fig.~\ref{suppl_fig6}B. 
For $L = 1$, most of the reconstructed images were blurred because the compressive sensing ratio is only 2.5 $\%$ at $T_N = 0.8$ ns, suggesting the shortage of information for the reconstruction. 
However, RMSE reduced as $L$ increased, and the reconstructed images were less blurred (Fig.~\ref{suppl_fig6}B). 
Incorporating WDM in the proposed approach is effective for increasing the acquisition rate while keeping the qualities of the reconstructed images. 

%
%
%

\end{document}